\documentclass[authoryear]{elsarticle}

\usepackage{soul}

\usepackage[english]{babel}
\usepackage{fancyhdr} 
\usepackage{url} 
\usepackage{subcaption} 
\usepackage{natbib} 
\usepackage{multirow} 
\usepackage{float}
\floatstyle{plaintop}
\restylefloat{table}

\usepackage{amssymb}
\usepackage[intlimits]{amsmath} 
\usepackage{amsthm} 
\usepackage[german]{nomencl}
\usepackage{booktabs} 
\usepackage{geometry}
\usepackage{enumitem}
\usepackage{multicol}

\makenomenclature
\usepackage[pdfborder={000},colorlinks=true,linkcolor=blue,citecolor=blue]{hyperref}
\usepackage{listings}
\usepackage[all]{xy}
\usepackage{color}

\makeatletter
\def\namedlabel#1#2{\begingroup
    #2%
    \def\@currentlabel{#2}%
    \phantomsection\label{#1}\endgroup
}
\makeatother

\definecolor{b}{rgb}{0,0,.8}	
\definecolor{g}{rgb}{0,.6,0}	
\definecolor{n}{rgb}{0,0,0}	
\definecolor{h}{rgb}{0.4,0.2,0.2}	
\definecolor{v}{rgb}{0.2,0.6,0}

\newcommand{\F}{{\mathbb F}}

\newcommand{\R}{{\mathbb R}}

\newcommand{\X}{{\mathbb X}}

\newcommand{\Z}{{\mathbb Z}}

\newcommand{\DD}{{\mathcal{D}}}
\newcommand{\EE}{{\mathcal{E}}}

\newcommand{\GG}{{\mathcal{G}}}

\newcommand{\JJ}{{\mathcal{J}}}
\newcommand{\KK}{{\mathcal{K}}}
\newcommand{\LL}{{\mathcal{L}}}

\newcommand{\OO}{{\mathcal{O}}}

\newcommand{\TT}{{\mathcal{T}}}

\newcommand{\XX}{{\mathcal{X}}}
\newcommand{\YY}{{\mathcal{Y}}}

\newcommand{\bsB}{\boldsymbol B}

\newcommand{\bsY}{\boldsymbol Y}

\newcommand{\bsnull}{\boldsymbol 0}

\newcommand{\bsbeta}{\boldsymbol \beta}

\newcommand{\bsxi}{\boldsymbol \xi}

\newcommand{\eps}{{\varepsilon}}

\DeclareMathOperator*{\argmin}{arg\,min}

\newcommand{\ov}\overline
\newcommand{\what}{\widehat}
\newcommand{\wtilde}{\widetilde}

\newcommand{\rig}\right
\newcommand{\lef}\left
\newcommand{\nf}\normalfont

\begin{document}

\title{Lasso Estimation for GEFCom2014 Probabilistic Electric Load Forecasting}
\author{Florian Ziel \corref{cor1}}
\ead{ziel@europa-uni.de}
\cortext[cor1]{Corresponding author}
\address{Europa-Universit\"at Viadrina, Frankfurt (Oder), Germany}

\author{Bidong Liu \corref{cor2}}
\ead{bliu8@uncc.edu}
\address{University of North Carolina at Charlotte, Charlotte, North Carolina, USA}

\journal{International Journal of Forecasting}

\begin{keyword}
Probabilistic forecasting \sep Threshold AR\sep Time-varying effects 
\end{keyword}
\begin{frontmatter}
\lhead{\nouppercase{\leftmark}}
\begin{abstract}

We present a lasso (least absolute shrinkage and selection operator) estimation based methodology for probabilistic load forecasting.
The considered model can be regarded as a bivariate time-varying threshold autoregressive(AR) process 
for the hourly electric load and temperature.
The joint modeling approach directly incorporates the temperature effects and reflects daily, weekly, and annual seasonal patterns and public holiday effects. 
We provide two empirical studies, one based on the probabilistic load forecasting track of the Global Energy Forecasting Competition 2014 (GEFCom2014-L), 
and the other based on another recent  probabilistic load forecasting competition that follows the similar setup as GEFCom2014-L.
In both empirical case studies, the proposed methodology outperforms two multiple linear regression based benchmarks from a top 8 entry of GEFCom2014-L.

\end{abstract}
\end{frontmatter}

\section{Introduction}

We present a lasso (least absolute shrinkage and selection operator) estimation based methodology for probabilistic load forecasting.
The lasso estimator introduced by \cite{tibshirani1996regression} has the properties of  automatically shrinking parameters and selecting variables. It thus enables us to estimate high-dimensional parameterizations. 
The procedure learns from the data in the sense that the parameters of less important variables will automatically get minor or even zero value.
The considered time series model is a bivariate time-varying threshold autoregressive (AR) model for hourly load and temperature. 
The model is specified so that it captures several stylized facts in load forecasting, such as the underlying daily, weekly, and annual seasonal patterns, the non-linear relationship between load and temperature, and holiday and long term effects.

In this paper, we illustrate the proposed methodology using two case studies from two recent forecasting competitions. 
The first one was from the probabilistic load forecasting track of Global Energy Forecasting Competition 2014, denoted as GEFCom2014-L. 
The topic of  GEFCom2014-L is month-ahead hourly probabilistic load forecasting with hourly temperature from 25 weather stations.
More details about GEFCom2014-L such as rules and data can be found in \cite{hong2015probabilistic}.
When implementing the proposed methodology, we create a new virtual temperature time series by averaging the temperature of stations 3 and 9.
These stations are chosen, as they give the best in-sample fit with
a cubic regression of the load against the temperature.

The second one was from the year-ahead probabilistic load forecasting competition organized by Tao Hong from UNC Charlotte in fall 2015, which was an extended version of GEFCom2014-L. In this paper we refer this competition as GEFCom2014-E. 
The competition included 5 tasks.
In each task, the participants were asked to forecast the next year of hourly load and submit the forecasts in 99 quantiles.
6 years (2004-2009) of hourly temperature and 4 years (2006-2009) of hourly load data was provided as the historical data for the first task.
In each of the remaining 4 tasks, an additional year of hourly load and temperature for the forecasted period of the previous task was provided.
The data for GEFCom2014-E can also be found in \cite{hong2015probabilistic}.
 Florian Ziel joined this competition with the proposed methodology, ranking top 2 out of 16 participating teams.

The structure of this paper is as follows:
in section 2,  we introduce the time series model; in section 3, we discuss the lasso estimation algorithm;  
in section 4, we describe two benchmarks developed from the methodology used by Bidong Liu to win a top 8 place in GEFCom2014-L; and in section 5, we present the empirical results. The paper is concluded in section 6.

.

\section{Time Series Model}

Let $(\bsY_t)_{t\in \Z}$ with $\bsY_t = (Y_{\LL,t}, Y_{\TT,t})'$ be the $d=2$-dimensional time series model of interest 
and denote $\DD = \{\LL,\TT\}$. So that $Y_{\LL,t}$ is the electric load and $Y_{\TT,t}$ the temperature 
at time point $t$.

The considered joint multivariate time-varying threshold AR model (VAR) for $(\bsY_t)_{t\in \Z}$ is given by
\begin{equation}
 Y_{i,t} = \phi_{i,0}(t) + \sum_{j \in \DD} \sum_{c \in C_{i,j}} \sum_{k\in I_{i,j,c} } 
 \phi_{i,j,c,k}(t) \max\{ Y_{j,t-k}, c\} +  \eps_{i,t}
 \label{eq_main_ar_model}
\end{equation}
for $i \in \DD$ where $\phi_{i,0}$ are the time-varying intercepts and $\phi_{i,j,k,c}$ are time-varying autoregressive coefficients. 
Moreover, $C_{i,j}$ are the sets of all considered thresholds, $I_{i,j,c}$ are the index sets of the corresponding lags and $\eps_{i,t}$ is the error term. We assume that the error process is uncorrelated with zero mean and constant variance.

Furthermore,
it is important that we are using the whole dataset with all hours to model hourly load and temperature, instead of
using dataset sliced by hour to model load with a specific hour as often done in literature. Forecasting algorithms
applied on the whole dataset can learn better about those events since the full dataset is more informative
than each small hourly dataset.

The modeling process has three crucial components:
The choice of the thresholds sets $C_{i,j}$, the choice of the lag sets $I_{i,j,k}$ and 
the time-varying structure of the coefficient. We describe these issues in the following three subsections.

\subsection{Choice of the threshold sets}

The choice of the thresholds sets $C_{i,j}$ will characterize the potential non-linear impacts in the model.
Note that if we choose $C_{i,j} = \{-\infty\}$ model  \eqref{eq_main_ar_model} will turn into a standard multivariate
time-varying AR process.

\begin{figure}[hbt!]
\centering
\begin{subfigure}[b]{0.49\textwidth}
 \includegraphics[width=1\textwidth]{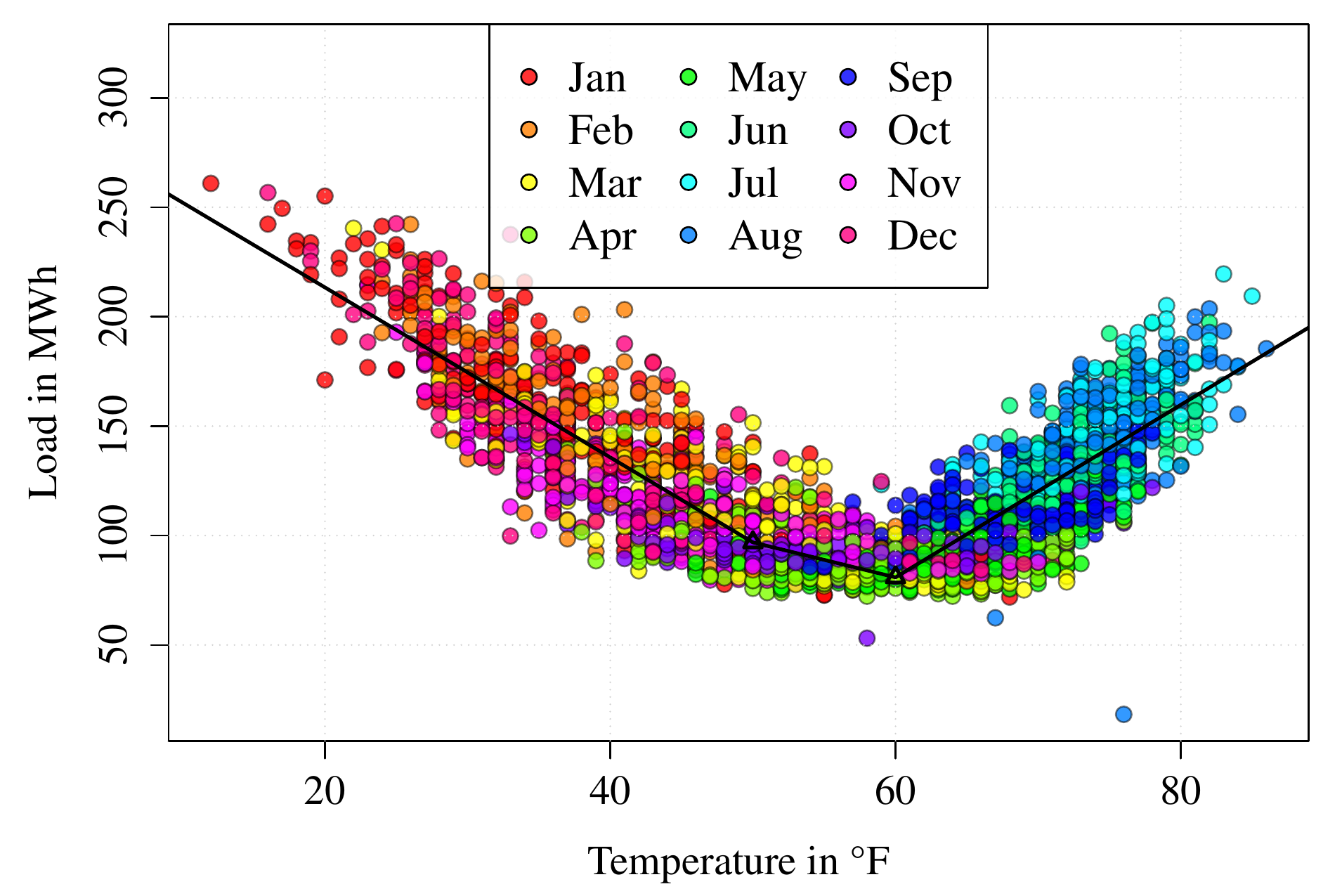}
   \caption{GEFCom2014-L data}
    \label{fig_wt_sub1}
\end{subfigure}
\begin{subfigure}[b]{0.49\textwidth}
 \includegraphics[width=1\textwidth]{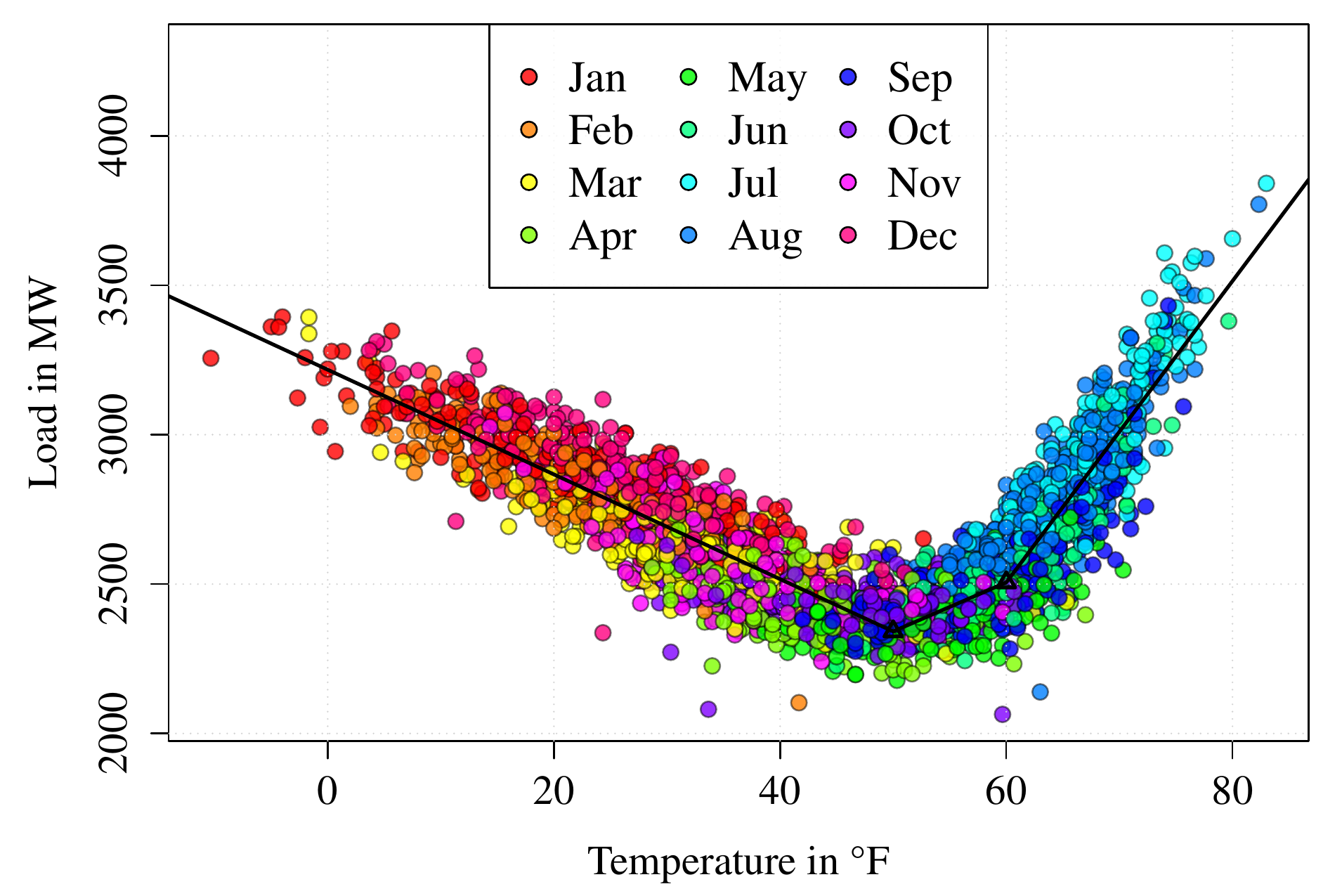}
   \caption{GEFCom2014-E data}
  \label{fig_wt_sub2}
\end{subfigure}
\caption{Temperature against load for all days at 00:00 with fitted values of model \eqref{eq_mod_example} for both data sets.
}
 \label{fig_tempthresh_motivation}
\end{figure}

For load data there is typically a non-linear effect of the temperature to the electric load.
Figure \ref{fig_tempthresh_motivation} shows the temperature of every day at 00:00 in the sample against the corresponding load.
We observe in general a decreasing relationship in the lower temperature area and an increasing one 
for larger degrees. 
To emphasize the non-linear relationship, we added the fitted line of the toy example regression 
\begin{equation}
 Y_{\LL,t} =  c_0 + c_1 Y_{\TT,t} + c_2 \max\{Y_{\TT,t} , 50\} + c_3 \max\{Y_{\TT,t} , 60\} + \epsilon_t .
 \label{eq_mod_example}
\end{equation} 
This is a simple threshold model with thresholds at 50$^\circ$F and 60$^\circ$F.

In Figure  \ref{fig_tempthresh_motivation} we see that the threshold model \eqref{eq_mod_example}
captures the relationship by piecewise linear functions.
Even though this is just an illustrative example, we see that this type of model is able to approximate 
all non-linear relationships between load and temperature. 

We can also introduce many other thresholds in the model to obtain more flexibility. However, it will enlarge the parameter space, which
brings with longer computation time and raises the concern of over-fitting. The lasso estimation algorithm can help to
ease these two concerns. Even better, it will only keep significant non-linear impacts.

For both data sets we choose the threshold sets manually. 
For the GEFCom2014-L data, we consider
$C_{\LL,\TT} =  \{ -\infty, 20, 30, 40, 45, 50, 55, 60, 65, 70, 80 \}$ 
 for thresholds of the temperature to electric load impact 
and $C_{\LL,\LL} =  \{ -\infty, 100, 125, 150, 175, 200, 225\}$ for the load to load effects. 
Remember that the thresholds corresponding with $-\infty$ model the linear effects.
For the other sets we assume no non-linear effects, so  $C_{\TT,\LL} = C_{\TT,\TT} = \{-\infty\}$.
For the GEFCom2014-E data we are using different thresholds, as the data is on a different scale.
In detail we use 
$C_{\LL,\TT} =  \{ -\infty, 10, 20, 30, 40, 45, 50, 60, 70, 80 \}$,
$C_{\LL,\LL} =  \{ -\infty, 2500, 3000, 3500, 4000, 4500\}$ and $C_{\TT,\LL} = C_{\TT,\TT} = \{-\infty\}$ for the thresholds sets.
Note that in general a data driven threshold set selection is plausible as well, e.g. by a set of selected quantiles.

\subsection{Choice of the relevant lag sets}

The lag sets $I_{i,j, c}$ are essential for a good model as they characterize 
the causal structure of the processes and the potential memory of the process.
The lags in $I_{i,j, c}$ describe a potential lagged impact of regressor $j$ at threshold $c$ to the process $i$.
It is widely known that the load at time $t$ is related to both its past and
the temperature. 
Therefore we choose $I_{\LL,\LL, c}$ and $I_{\LL, \TT, c}$ non-empty for all $c$.
For the temperature the situation is slightly different. Here, we assume that the temperature
depends on its past, so $I_{\TT,\TT, -\infty}$ is non-empty as well. 
But, it is clear that the electric load
does not effect the temperature, so $I_{\TT, \LL, -\infty}$ is empty.

The selected index sets are given in Table \ref{tab_lags}.
\begin{table}[tbh]
\centering
\begin{tabular}{ll}
 Index sets & Contained Lags  \\ 
  \hline \hline
$I_{\LL, \LL, -\infty}$  & $1, \ldots, 1200$ \\ \hline 
$I_{\LL, \LL, c}$ (with $c\neq -\infty$), $I_{\LL, \TT, c}$  & $1, \ldots, 200$ \\ \hline 
$I_{\TT, \TT, -\infty}$  & $1, \ldots, 360$ \\ \hline 
$I_{\TT, \LL, -\infty}$  & -  
\end{tabular}
\caption{ Considered lags of the required index sets.}
\label{tab_lags}
\end{table}
Here, similarly as for the threshold sets, larger sets increase the parameter space and consequently result in computational burden. 
Still they have to be chosen large enough to capture the relevant information. 
$I_{\LL, \LL, -\infty}$ contains all lags up to 1200, so the maximal memory is the preceding 1200 hours, slightly more than 7 weeks.
The most essential part is that the important lags of orders such as 1, 24, 48 and 168 are included.
A detailed discussion for the choice of the index sets 
can be found in \cite{ziel2015efficient}.

\subsection{The time-varying coefficients}

The assumed structure of the time-varying coefficients is substantial as well. 
They have big impacts not only
on seasonality and public holiday effects, but also on the long term trend behavior.
Still, we keep most of the coefficients constant, allowing only the important ones to vary over time.
The intercepts $\phi_{i,0}$ in equation \eqref{eq_main_ar_model} are important and are allowed to vary over time 
for both the load and the temperature.
For the load we additionally allow for $\phi_{\LL,\LL,-\infty,k}$ with $k\in \{1,2, 24,25\}$ to vary over time, 
and for the temperature $\phi_{\TT,\TT,-\infty,k}$ with $k\in \{1,2\}$. 
So in total the 2 intercepts, 4 autoregressive load, and 2 autoregressive temperature coefficients are allowed to vary over time.
Obviously, this choice can be modified based on knowledge about the important parameters.
And again, it holds true that the more parameters vary over time, the larger the parameter space. Thus, the computation time increases and 
the limited over-fitting risk as well. 

For the time varying coefficients we assume a similar structure as in \cite{ziel2015efficient}.
We assume for a time-varying parameter of interest $\xi$ (e.g. $\phi_{i,0}$ or $\phi_{i,j,c,k}$) that
\begin{equation}
  \xi(t)  = \xi_0 + \bsxi' \bsB^{\xi}(t) = \xi_0 + \sum_{l=1}^{N_\xi} \xi_l B^{\xi}_l(t) 
  \label{eq_basis}
\end{equation}
where $\bsxi= (\xi_1, \ldots, \xi_{N_\xi})'$ is the vector of coefficients that applies to the basis functions 
$\bsB^{\xi} = (B^{\xi}_1, \ldots, B^{\xi}_{N_\xi})'$.
Obviously, the sum in \eqref{eq_basis} is empty for constant parameters.

The basis functions of the time-varying coefficients have to be chosen accurately.
The selection is modular. 
Several effects can be added and merged easily.
We consider a selection of several groups of regressors as listed in Table \ref{tab_groups}.
\begin{table}[tbh]
\centering
\begin{tabular}{ll}
 Group & Description  \\ 
  \hline \hline
$\GG_1$ & hourly impacts on the seasonal daily pattern \\ \hline 
$\GG_2$ & hourly impacts on the seasonal weekly pattern \\ \hline 
$\GG_3$ & daily impacts on the seasonal annual pattern \\ \hline 
$\GG_4$ & smooth annual impacts \\ \hline 
$\GG_5$ & long term trend effects \\ \hline 
$\GG_6$ & fixed date public holidays effects\\ \hline
$\GG_7$ & varying date public holidays effects\\ \hline
$\GG_8$ & interaction effects between $\GG_1$ and $\GG_4$ 
\end{tabular}
\caption{List of all considered groups $\GG_1, \ldots, \GG_8$ of basis functions}
\label{tab_groups}
\end{table}

%

\begin{figure}[hbt!]
\begin{subfigure}[b]{0.49\textwidth}
 \includegraphics[width=1\textwidth]{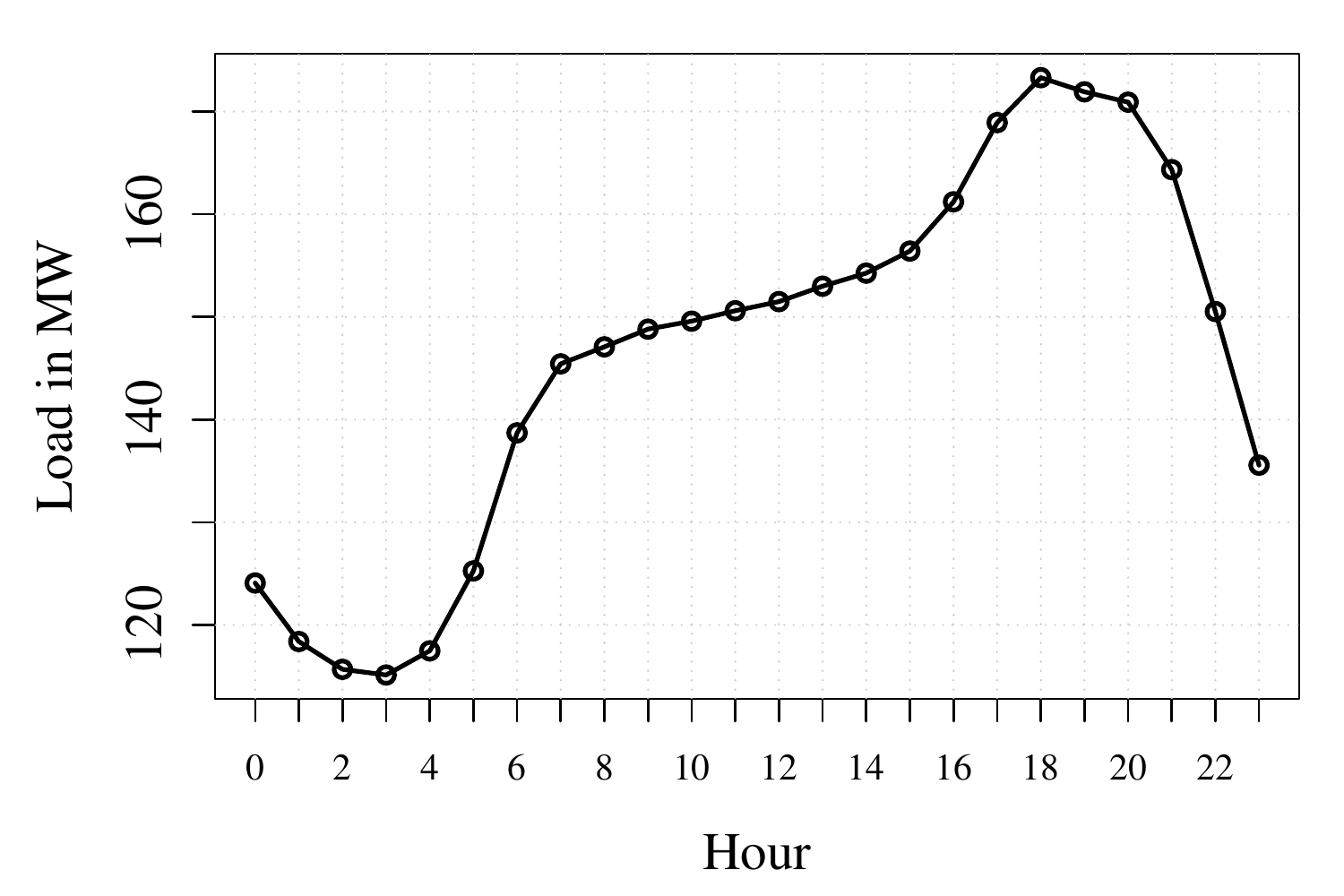}
   \caption{Hourly mean load during a day}
  \label{fig_w_sub1}
\end{subfigure}
\begin{subfigure}[b]{0.49\textwidth}
 \includegraphics[width=1\textwidth]{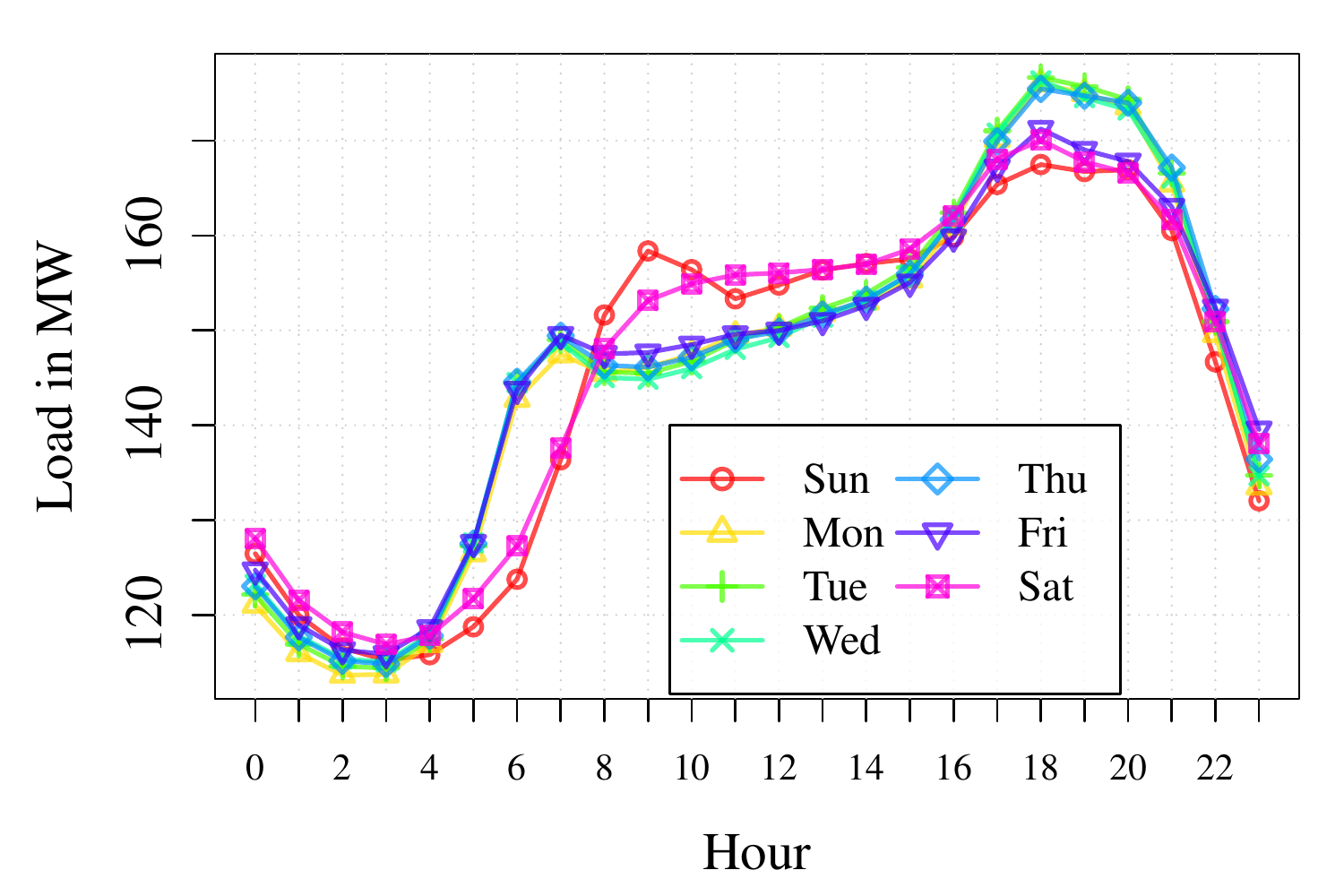}
   \caption{Hourly mean load during a week}
  \label{fig_w_sub2}
\end{subfigure}
\caption{Hourly mean load during a day (\ref{fig_w_sub1}) and week (\ref{fig_w_sub2}) of the GEFCom2014-L data }
 \label{fig_w}
\end{figure}

Below we explain the groups $\GG_1, \ldots, \GG_8$ one by one. 
The daily and the weekly mean electric load of the GEFCom2014-L data is given in Figure \ref{fig_w}.
In \ref{fig_w_sub1} we see the clear distinct seasonal daily pattern, with low values during night and high values during the day.
The group $\GG_1$ will cover this effect. Obviously, this requires 24 parameters. 
However, in \ref{fig_w_sub2} we observe that the Saturdays and Sundays 
show different behaviors to the typical working days from Monday to Friday,
which exhibit basically the same behavior every day. 
Nevertheless, there is a transition effect on Monday morning and Friday evening towards and 
from the weekend. 
$\GG_2$ will cover the full weekly structure and 168 parameters are required.
As mentioned, there is redundancy in the pattern, e.g. the Tuesdays, Wednesdays and Thursdays generally exhibit similar behaviors.
This structure is automatically taken into account when using the regressors $\GG_1$ and $\GG_2$ in combination with 
the lasso estimation technique. 
The basis function of 
group $\GG_1$ and $\GG_2$ are defined by
\begin{align}
B_k^{\GG_1}(t) = \begin{cases}
                  1 & , k \leq \text{HoD}(t) \\
                  0 & , \text{otherwise} \\
                 \end{cases}
\ \ \  \text{ and } \ \ \
B_k^{\GG_2}(t) = \begin{cases}
                  1 & , k \leq \text{HoW}(t)  \\
                  0 & , \text{otherwise} \\
                 \end{cases}     
\label{eq_hod_wod}
\end{align}
where $\text{HoD}(t)$ and $\text{HoW}(t)$ gives the hour-of-the-day ($1, 2, \ldots, 24$) and 
the hour-of-the-week ($1, 2, \ldots, 168$, start counting at Sunday 0:00) of time point $t$. 
Note that in \eqref{eq_hod_wod} the parametrization is done by 
cumulative components. Therefore the "$\leq$" relation is used instead of the commonly used "$=$" relation. 
As an example, $B_2^{\GG_1}$ models the additional impact of hour 1:00 to hour 0:00, which is modeled by $B_1^{\GG_1}$; 
instead of modeling the direct impact of hour 1:00 which would be associated with the  "$=$" relation in \eqref{eq_hod_wod}.
In other words, we are modeling the changes of the impacts associated with an hour, instead of the absolute effects.
Our estimation method will make a parameter included in the model only if the corresponding change is significant. 

Similarly to the daily and weekly pattern, there is an annual seasonal pattern. To capture this we introduce 
\begin{align}
B_k^{\GG_3}(t) = \begin{cases}
                  1 & , k \leq \text{DoY}(t) \\
                  0 & , \text{otherwise} \\
                 \end{cases}
\end{align}
where $\text{DoY}(t)$ gives the day-of-the-year ($1, 2, \ldots, 365$) of time point $t$ in a common year with 365 days. 
In a leap year $\text{DoY}(t)$ also takes values from ($1, 2, \ldots, 365$), but the 
29th February has the same value (namely 59) as the 28th February. Similarly as above, we model the changes in the annual pattern,
not the direct impact. 

\begin{figure}[hbt!]
\begin{subfigure}[b]{0.49\textwidth}
 \includegraphics[width=1\textwidth]{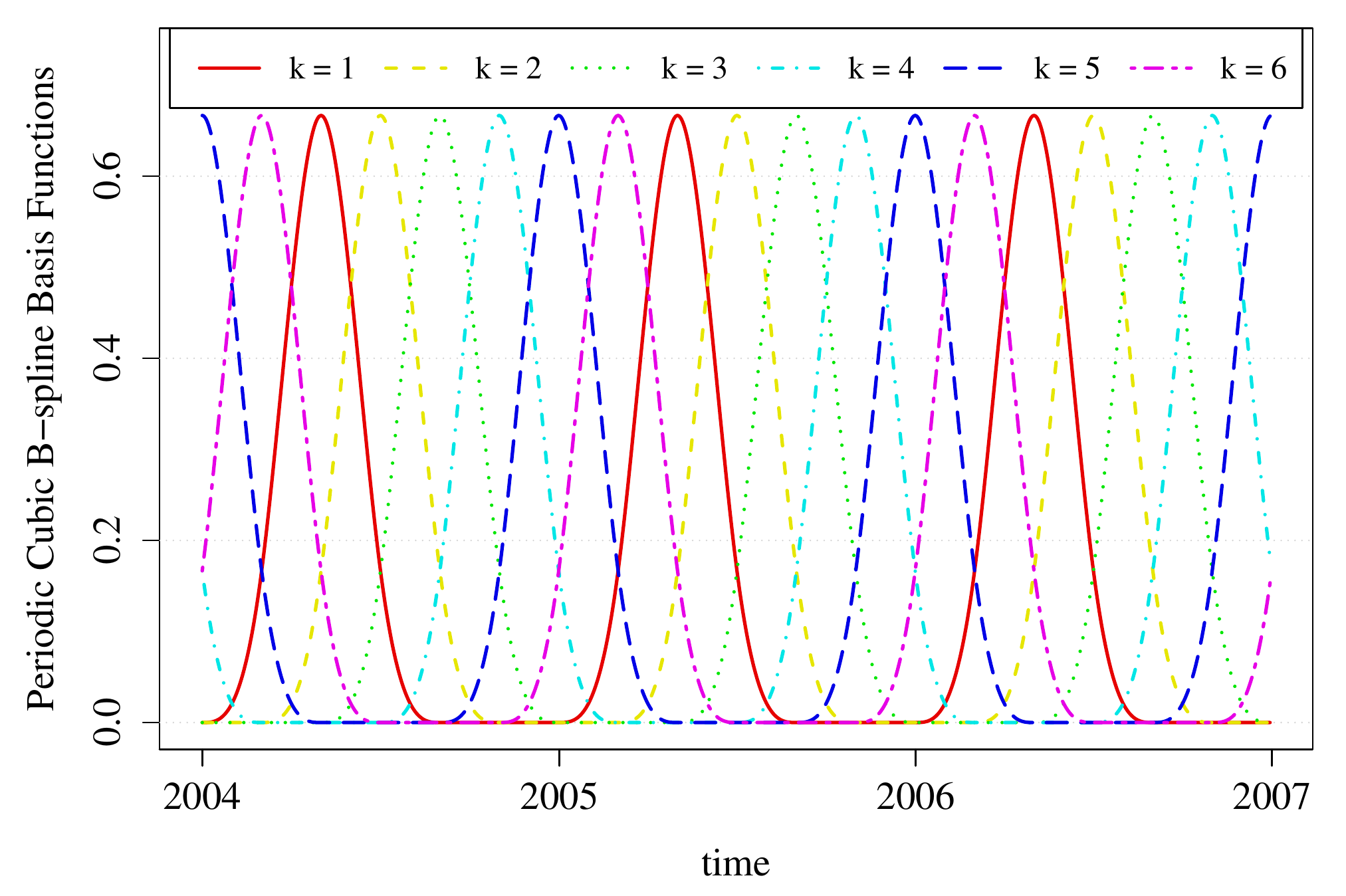}
   \caption{Periodic cubic B-spline basis within 3 years}
  \label{fig_wx_sub1}
\end{subfigure}
\begin{subfigure}[b]{0.49\textwidth}
 \includegraphics[width=1\textwidth]{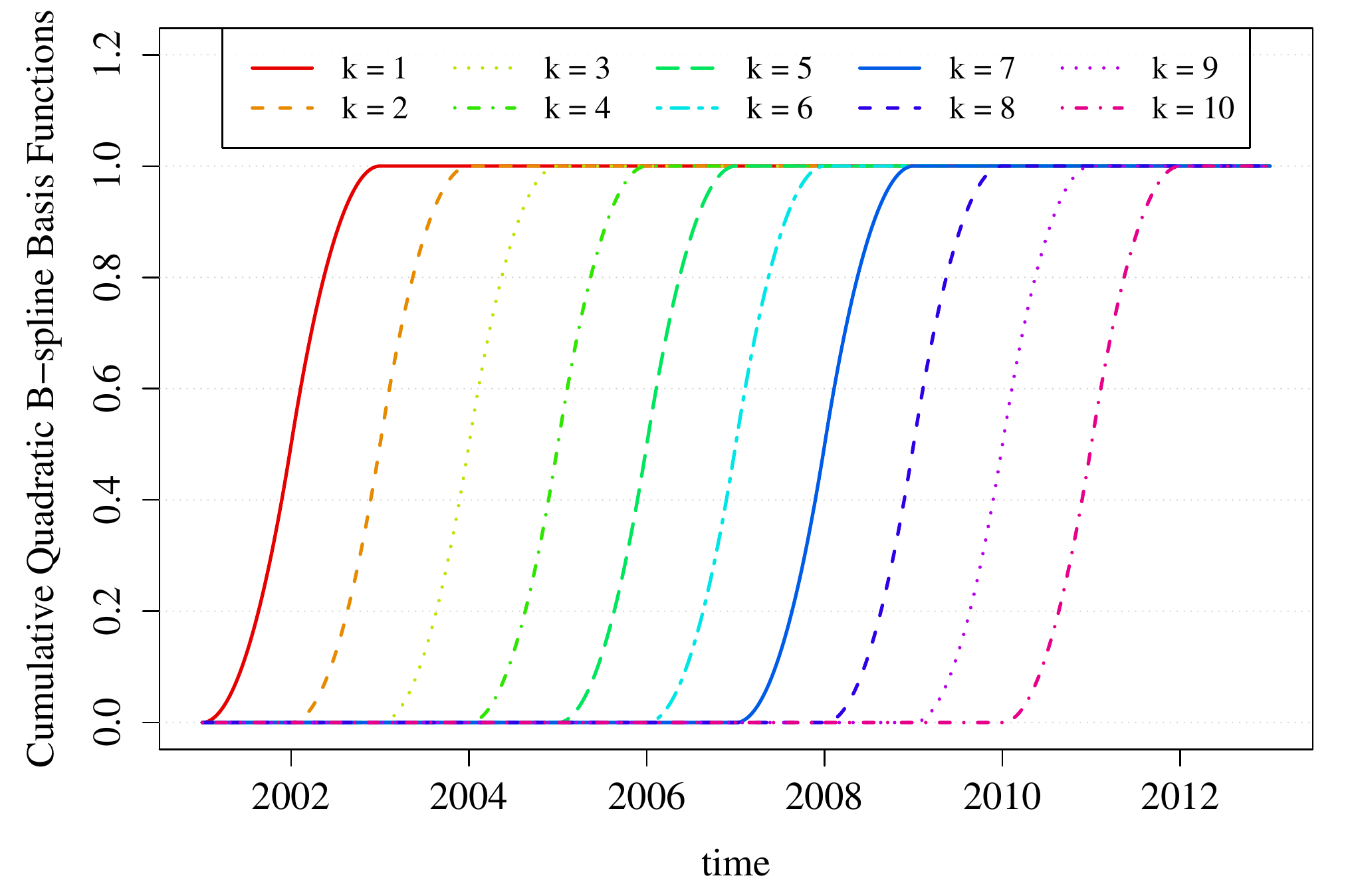}
   \caption{Cumulated quadratic B-spline basis on 12 years}
  \label{fig_wx_sub2}
\end{subfigure}
\caption{Illustration of basis functions for $\GG_4$ and $\GG_5$}
 \label{fig_wx}
\end{figure}

The next group of basis functions concerns smooth annual impacts. 
This will capture similar effects as in $B_k^{\GG_3}$ but more in a smooth manner. 
We consider periodic B-splines which results in a local modeling approach.
In detail, we use cubic B-splines with a periodicity of $8765.76 = 24 \times 365.24 $ on an equidistant grid with 6 basis functions.
In graph \ref{fig_wx_sub1} we see these basis functions on a time range of three years. We clearly observe 
the local impact. So e.g. the dashed yellowish function ($k=2$) covers only effects in the summer, but has no impact in the winter.

The most tricky basis function group concerns the long term effects. 
The challenging part is the distinction between spurious effects and real long term changes in the load behavior.
The spurious effect problem is crucial for long term forecasting, whereas for shorter time horizons it is negligible.
To make the problem clear, suppose the available time series ends in 31th December. Suppose the last two months, November and
December, had low load values for some unknown reason. Now the question is, if this was a random effect (just a realization of rare or outlier events) or a structural change in the load level (induced e.g. by better energy efficiency which is not captured by external regressors). The conservative way of statistical modeling would suggest a random effect, unless the structural change
is significant enough to be detected by the modeling approach. 

We model long term effects by monotonically increasing basis functions. 
They are constant in the past, then strictly monotonically increasing in a certain time range where the long term 
transition effect might have taken place, then constant after this possible transition. 
The time range where the basis function is monotonically increasing should be larger than a year to reduce the probability to include spurious effects.
Furthermore, the distance between these basis functions should be relatively large as well. 
We consider a distance of one year between the basis functions with a support of two years for the transition effect.
In detail, we use cumulative quadratic B-splines as basis function for the long term effects.
We consider only basis functions where the in-sample basis functions take a smallest value of at least $10\%$ of the overall maximum and at most  $90\%$ of the overall maximum.
This will reduce the danger of modeling a spurious effect.
We end up with only a few basis functions. 
An illustrative example for an in-sample period of 12 years (2001 to 2012) with 
the out-of-sample year 2013 is given in Figure \ref{fig_wx_sub2}. 
Note that the number of the long term basis functions in group $\GG_5$ depends on the data range.

The next two groups $\GG_6$ and $\GG_7$ contain the public holiday information.
In general electric load exhibits a special behavior at public holidays, 
that eventually disturbs the standard weekly pattern. 
For modeling purpose, we group the public holidays into two classes: with fixed date such as New Year's Day (Jan. 1) and with a flexible date such as Thanksgiving Day (fourth Thursday in Nov.). 
We consider all United States federal public holidays. 
We denote the sets of public holidays with fixed and
flexible date by $\F\text{ix}$ and $\F\text{lex}$.

As days in
$\F\text{lex}$ are always at a specific weekday, we can expect the same behavior every
year at these public holidays.
If there is a week with a public holiday, then the typical weekly structure in Figure \eqref{fig_w_sub2} changes. 
Not only the structure of the public holiday is affected,
but also the hours before and after the public holiday, due to transition effects. 
Therefore we define for each flexible public holiday $F \in \F\text{lex}$ a basis of $6+24+6 = 36$ hours 
(6 hours before $F$, 24 hours at $F$, and 6 hours after $F$).
In detail, it is given by 
$$ B^{F}_k(t) = \begin{cases}
                  1 & , k \leq \text{Ho}F(t) \\
                  0 & , \text{otherwise} \\
                 \end{cases} $$
where Ho$F(t)$ gives the hours from $1,2, \ldots, 36$ at time point $t$ around the public holiday starting counting from 18:00.

The impact of the days in
$\F\text{ix}$ is complex, because it depends on the weekday of incidence.
Some research found it is usually similar to that of a Sunday (see e.g. \cite{ziel2015efficient}). 
We will introduce an effective coefficients $C(t)$ for each hour of the week. 
With $C(t)$ we can define the basis functions for $H \in \F\text{ix}$
$$ B^{H}_k(t) = \begin{cases}
                  C(t) & , k \leq \text{Ho}H(t)\\
                  0 & , \text{otherwise} \\
                 \end{cases}, $$
where Ho$H(t)$ gives the hours from $1,2, \ldots, 36$ at time point $t$ around the public holidays starting counting from 18:00.
The coefficients $C(t)$ are defined as follows: 
If the public holiday is on a Sunday, then the effective coefficient is 0, assuming that there 
is no additional impact of the public holiday on a Sunday. Thus, we call these 24 hourly mean load values as low level load target.
If such a public holiday occurs during the core working days such as Tuesday, Wednesday or Thursday, we expect a full impact with the effective coefficient of 1.
We call the 24 hourly mean load values of these three days 
as high level load target. 
If the holiday happens on Monday, Friday or Saturday, the impact then should be between above two situations and the effective coefficient is usually between 0 and 1. 
If we denote the hourly mean load of the week from Figure \ref{fig_w_sub2} by actual load target, 
then we define the coefficients by
$ C(t) = \max\{ 1 - \frac{\text{high level load target}(t) - \text{actual load target}(t)}{ \text{high level load target}(t) - \text{low level load target}(t) }   , 1\}$.

The last group of basis functions focuses on interaction effects, which is important for the temperature modeling.
As the length of the night is changing over the year,
the daily seasonal pattern change over the year as well. 
We create the interaction group by multiplying each basis function of one group 
with the basis function of another group. Thus, the interaction groups tend to require many parameters.
For that reason we consider for the last group $\GG_8$ only the multiplication of the daily seasonal component $\GG_1$ with 
the smooth annual basis functions $\GG_4$. In detail, $\GG_8$ contains the basis functions
$ B^{\GG_8}_{24(j-1)+i}(t) = B_i^{\GG_1}(t)  B_j^{\GG_4}(t) $
for  $i\in \{1,\ldots, 24\}$ and $j \in \{1, \ldots, 6\}$.


With all basis function groups, we can define the full basis function vector $\bsB^{\xi}$ for 
a parameter $\xi$.
Hence, the basis functions for a time-varying parameter $\xi_\LL$ associated with the load is given by 
$\bsB^{\xi_\LL} = (\bsB^{\GG_1}, \bsB^{\GG_2}, \bsB^{\GG_3}, \ldots, \bsB^{\GG_8})$
where $ \bsB^{\GG_1} = (B_1^{\GG_1}, \ldots, B_{24}^{\GG_1} ) $,
$ \bsB^{\GG_2} = (B_1^{\GG_2}, \ldots, B_{168}^{\GG_2} ) $,
$ \bsB^{\GG_3} = (B_1^{\GG_3}, \ldots, B_{365}^{\GG_3} ) $, $ \bsB^{\GG_4} = (B_1^{\GG_4}, \ldots, B_{6}^{\GG_4} ) $, $\ldots$
define the vectors of the basis functions. For the time-varying parameters $\xi_\TT$ of  the 
temperature modeling process, we define $\bsB^{\xi_\TT} = (\bsB^{\GG_1}, \bsB^{\GG_4}, \bsB^{\GG_8})$.
Thus, only daily, smooth annual, and their interaction effects are allowed. 
Especially, we do not include any weekly, public holiday or long term effects for modeling the temperature.

\section{Estimation and Forecasting Method}

In the introduction we mention that we use a lasso estimation technique which is 
a penalized ordinary least square regression estimator.
The ordinary least square (OLS) representation of \eqref{eq_main_ar_model} is given by
\begin{equation}
 \YY_i = \XX_i \bsbeta_i + \EE_i .
\label{eq_ar_ols} 
\end{equation}
Here we denote
$\YY_i = (Y_{i,1}, \ldots, Y_{i,n})'$, $\XX_i$ the $n\times p_i$-dimensional regressor matrix that corresponds to \eqref{eq_main_ar_model}, 
$\bsbeta_i$ the full parameter vector of length $p_i$, $\EE_i= (\eps_{i,1}, \ldots, \eps_{i,n})'$ the residual vector,
and $n$ as number of observations.
However, we do not perform a lasso estimation for \eqref{eq_ar_ols} directly, but for its standardized version.
Therefore we standardize \eqref{eq_ar_ols} so that the regressors and the regressand have all variance 1 and mean 0. Thus 
we receive the standardized version of \eqref{eq_ar_ols}:
\begin{equation}
 \wtilde{\YY}_i = \wtilde{\XX}_i \wtilde{\bsbeta}_i + \wtilde{\EE}_i .
\label{eq_ar_ols_scaled} 
\end{equation}
We can easily compute $\bsbeta_i$ by rescaling, if $\wtilde{\bsbeta}_i$ is determined.
The lasso optimization problem of \eqref{eq_ar_ols_scaled} is given by
\begin{eqnarray}
 \what{\wtilde{\bsbeta}}_i &=  \argmin_{\bsbeta \in \R^{p_i} } \| \wtilde{\YY}_i -
 \wtilde{\XX}_i \bsbeta \|^2_2  + \lambda_{i} \|\bsbeta\|_1 \label{eq_lasso_ar} 
\end{eqnarray}
with tuning parameters $\lambda_i$, and $\|\cdot\|_1$ and $\|\cdot\|_2$ as $\ell_1$- and $\ell_2$-norm. 
For $\lambda_i = 0$, \eqref{eq_lasso_ar} is the standard OLS problem. For huge 
$\lambda_i$ values, we have a huge penalty on the parameters and receive the estimator 
$ \what{\wtilde{\bsbeta}}_i = \bsnull = (0, \ldots, 0)'$, so no parameter is included in the model. 
In a moderate range of $\lambda_i$ values, we get different solutions.
It holds that the larger $\lambda_i$, the less parameters are included in the estimated model.

\begin{figure}[hbt!]
\begin{subfigure}[b]{0.49\textwidth}
 \includegraphics[width=1\textwidth]{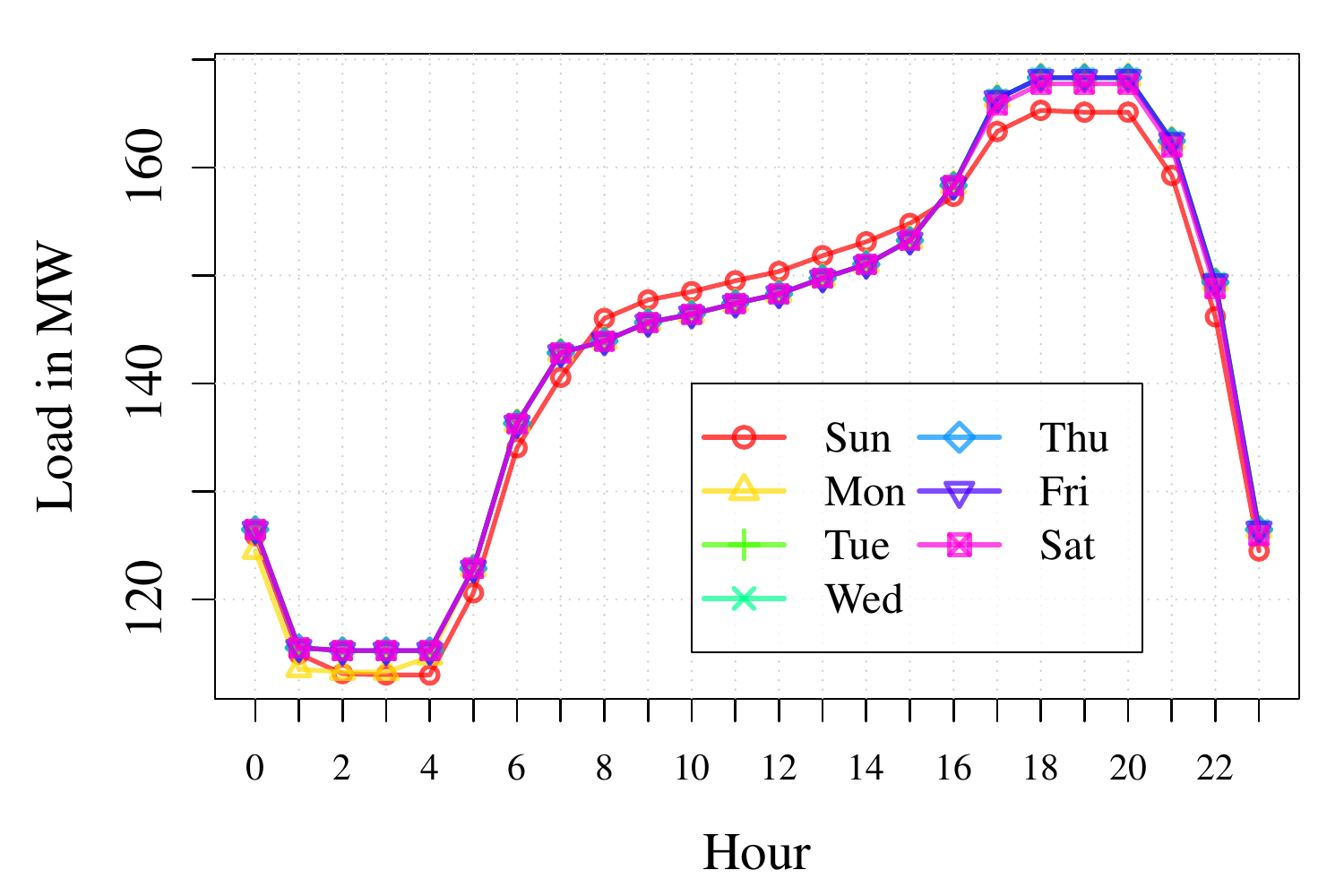}
   \caption{$\lambda=0.25$ with 28 non-zero parameters}
  \label{fig_wme_sub1}
\end{subfigure}
\begin{subfigure}[b]{0.49\textwidth}
 \includegraphics[width=1\textwidth]{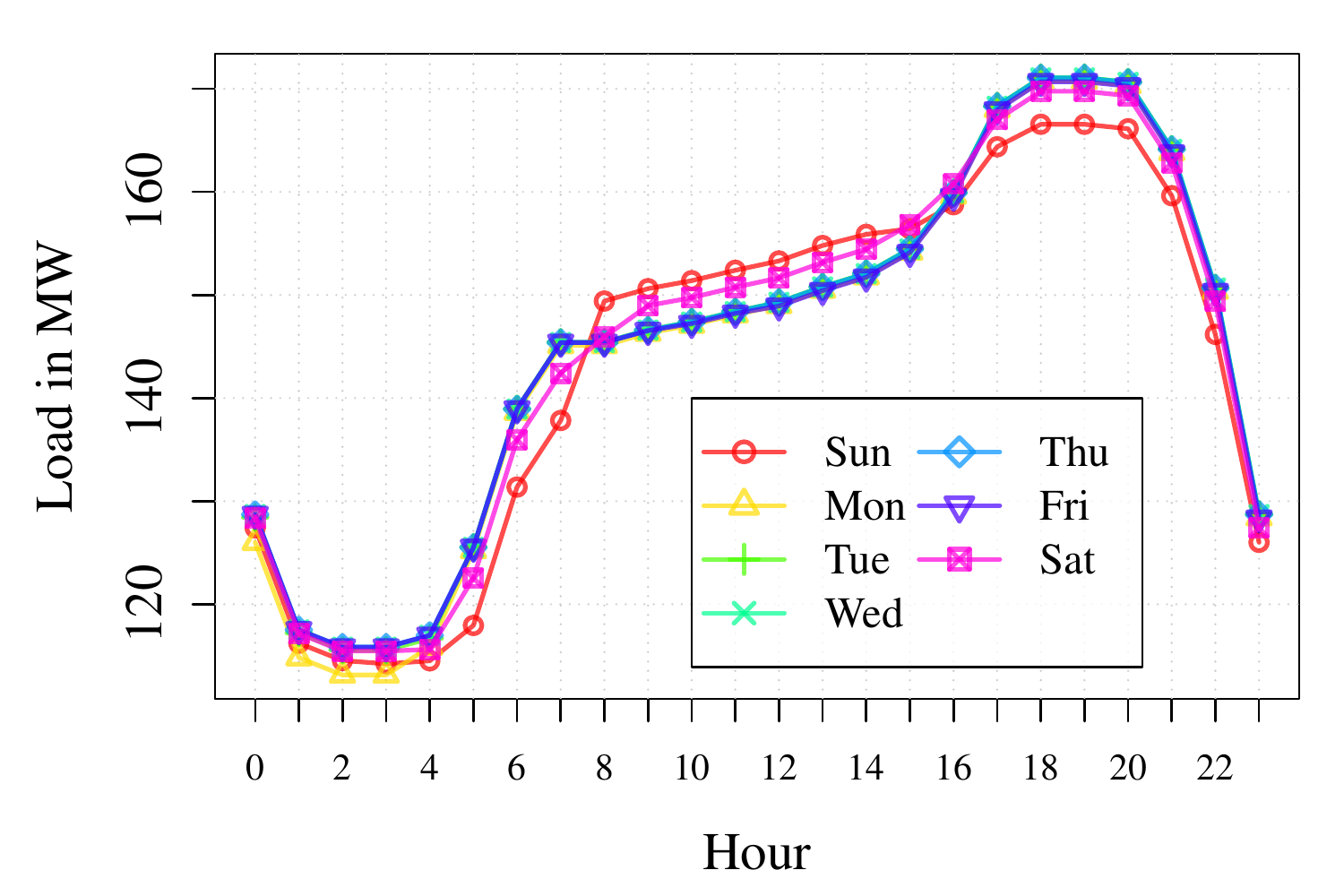}
   \caption{$\lambda=0.125$ with 43 non-zero parameters}
  \label{fig_wme_sub2}
\end{subfigure}
\begin{subfigure}[b]{0.49\textwidth}
 \includegraphics[width=1\textwidth]{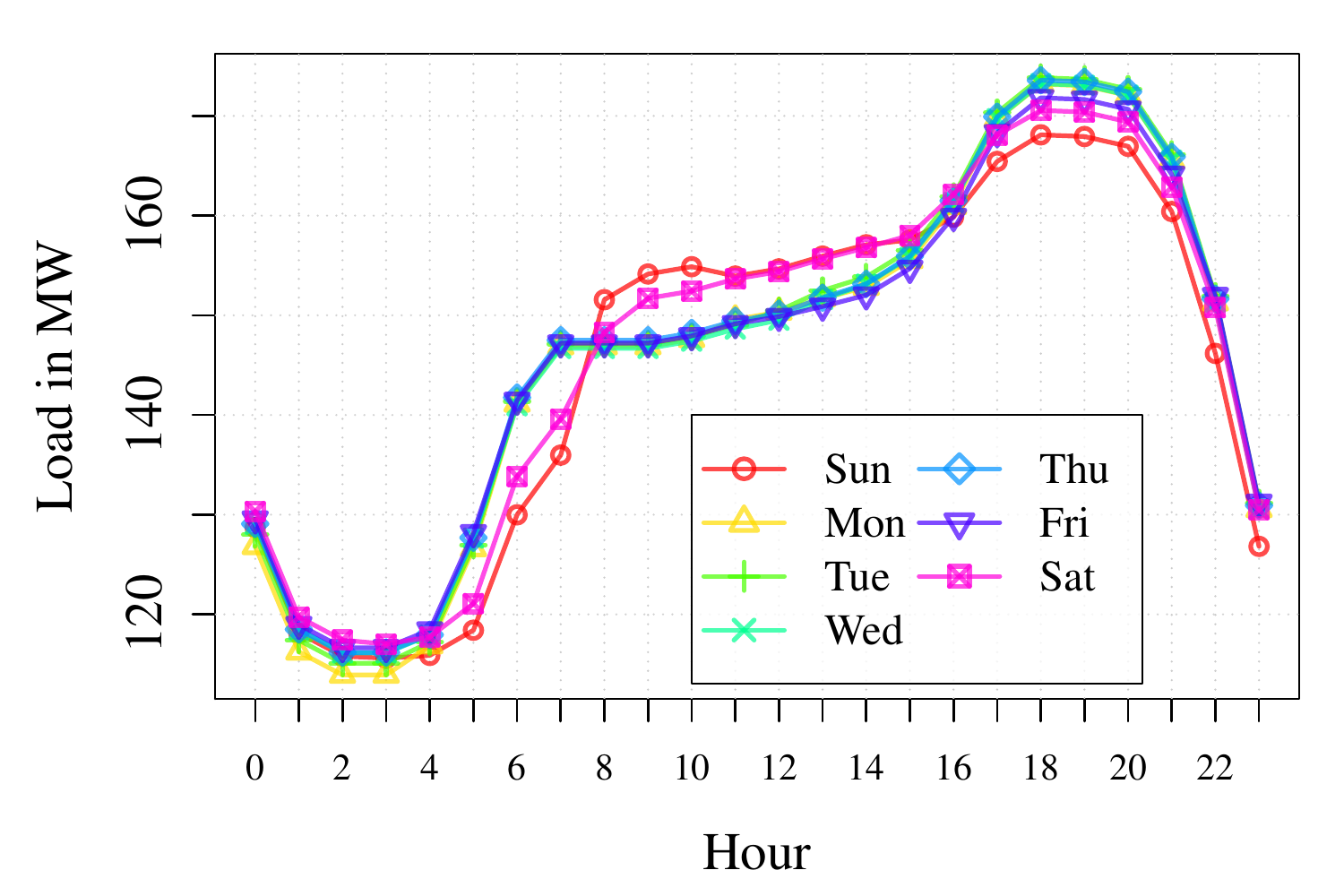}
   \caption{$\lambda=0.0625$ with 80 non-zero parameters}
  \label{fig_wme_sub3}
\end{subfigure}
\begin{subfigure}[b]{0.49\textwidth}
 \includegraphics[width=1\textwidth]{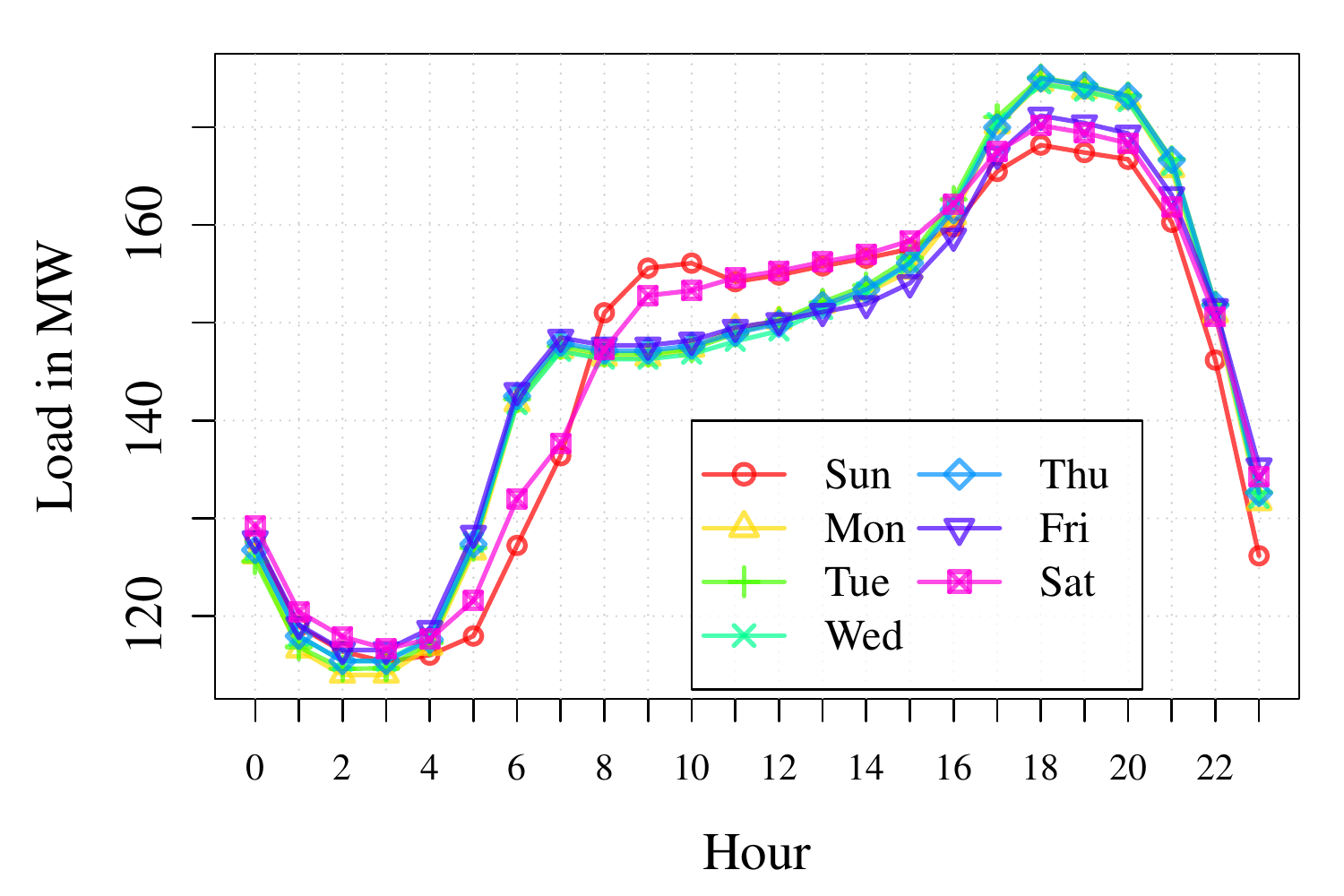}
   \caption{$\lambda=0.03125$ with 101 non-zero parameters}
  \label{fig_wme_sub4}
\end{subfigure}
\caption{Fitted model for model \eqref{eq_lasso_example} for selected $\lambda$ values with corresponding number of non-zero parameters.}
 \label{fig_wme}
\end{figure}

To better understand this feature, we consider a simple lasso problem given by
\begin{align}
 \| \YY_i - \X \bsbeta \|^2_2 + \lambda \|\bsbeta\|_1 ,
\label{eq_lasso_example}
 \end{align}
where $\X$ is the regressor matrix that contains the 24 basis functions of $\GG_1$ and the 168 basis functions of $\GG_2$.
We remember that the OLS solution of this problem corresponds to Figure \ref{fig_w_sub2} and requires 168 parameters
to full fully capture all effects.
In Figure \ref{fig_wme} we plot the fitted values of solution of \eqref{eq_lasso_example} for four different 
$\lambda$ values.
As mentioned, we see that the smaller $\lambda$, the more parameters are included in the model. 
Thus, the closer the solution gets to Figure in \ref{fig_w_sub2}.  
For example, in Figure \ref{fig_wme_sub3} we observe a pattern where the difference to Figure \ref{fig_w_sub2} 
is not easy to observe by eye-balling, even though only 80 parameters are required to capture the structure instead of 168.
In contrast \ref{fig_wme_sub1} with only 28 parameters does not cover the pattern well, so e.g. the 
seasonal pattern for all days except the Sunday is the same during the morning and noon hours. This indicates that 
the 28 parametric solution includes not enough parameters for an appropriate modeling.

Note that not only the selection property of the lasso is relevant, but also the shrinkage property.
For example, if we have the lasso solution in \ref{fig_w_sub2} with 80 non-zero parameters then this 
is different from the OLS solution of the corresponding 80 regressors. 
In general the lasso solution tends to have smaller estimated parameters (in terms of absolute values)
than the OLS solution, due to the shrinkage towards $\bsnull$.
In detail, the in-sample residual sums of square (RSS) is always larger for 
the lasso solution than for the OLS solution. 
Thus, even though there might be many non-zero parameters 
in the final estimated model, the contribution of many of the non-zero parameters to the model is small. 
This shrinkage property 
reduces the parameter uncertainty and might give better out-of-sample performance.

In general, the tuning parameters $\lambda_i$ should be chosen by a selection algorithm. 
Usually the optimal $\lambda_i$ will be chosen from a given grid $\Lambda_i$ by minimizing an information criterion. 
We select the tuning parameter with minimal Bayesian information criterion (BIC). 
The BIC is a conservative information criterion that avoids over-fitting.  
For the grid $\Lambda_i$ we choose an exponential grid as suggested by \cite{friedman2010regularization}.

As computation algorithm, we consider the fast coordinate descent algorithm 
and the corresponding \texttt{R} package functions of the \texttt{glmnet} package, see e.g. \cite{friedman2010regularization}
for more details.
The asymptotic computational complexity of the coordinate descent algorithm is only $\OO(np_i)$. 
This is optimal, as $np_i$ is the number of elements in the regression matrix. 
Thus, we can estimate the model efficiently and can easily carry out the model selection.
Another positive feature is that we do not require a division into training and test data set, as we 
can tune the model based on statistical theory (like the BIC).

For each forecasting task we use all available data for the lasso estimation procedure.
Given the estimated model, we can use residual based bootstrap to simulate future scenario sample paths as in \cite{ziel2015efficient}.
We consider in total $N=10000$ sample paths here.
The corresponding empirical percentiles are used as estimates for the target quantiles.

\section{Benchmarks}

The scenario-based probabilistic forecasting methodology proposed by \cite{hongjasonxie2014} was used by two top 8 teams (Jingrui Xie, top 3; Bidong Liu, top 8) in GEFCom2014-L.
In this paper, we develop two benchmarks using this method with two underlying models. 
The first one  is Tao\textsc{\char13}s Vanilla Benchmark  model used in GEFCom2012 \citep{hong2014global}, abbreviated as \textit{Vanilla} in this paper.
The second one is a recency effect model proposed by \cite{wangliuhong2015}, abbreviated as \textit{Recency} in this paper.
In the GEFCom2014-L case study, instead of performing weather station selection as discussed in \cite{hongwhitepu2015}, 
we create a temperature series by averaging the 25 weather stations to keep the benchmarks simple and easily reproducible. Note that this is different from how the temperature series is created when implementing the lasso based methodology as discussed in section 1.

\subsection{\textit{Vanilla} model}

The \textit{Vanilla} model for the load $Y_{\LL,t}$ is given as:
\begin{equation}
\label{eqtvb}
Y_{\LL,t} = \beta_0 + \beta_1 \text{MoY}(t) + \beta_2 \text{DoW}(t) + \beta_3 \text{HoD}(t) + \beta_4 \text{DoW}(t)\text{HoD}(t) 
+ f(Y_{\TT,t}) + \epsilon_t,
\end{equation}
  where 
$\beta_i$ are the regression coefficients, $\text{MoY}(t)$ gives the month-of-the-year ($1, \ldots, 12$) of time $t$, 
$\text{DoW}(t)$ gives the day-of-the-week ($1, \ldots, 7$ with $\text{Sunday}=1, \text{Monday}=2, \ldots$) of time $t$,  $\text{HoD}(t)$ gives the hour-of-the-day ($1,\ldots, 24$) of time $t$ as for equation \eqref{eq_hod_wod}
and
\begin{align}
\label{eq:SisterModel:2}
f(Y_{\TT,t}) = \beta_5 Y_{\TT,t} &+ \beta_6 Y_{\TT,t}^2 + \beta_7 Y_{\TT,t}^3+ \beta_8 Y_{\TT,t} \text{MoY}(t) + \beta_9 Y_{\TT,t}^2 \text{MoY}(t)  \nonumber \\
&+ \beta_{10} Y_{\TT,t}^3 \text{MoY}(t) + \beta_{11} Y_{\TT,t} \text{HoD}(t) + \beta_{12} Y_{\TT,t}^2 \text{HoD}(t) + \beta_{13} Y_{\TT,t}^3 \text{HoD}(t).
\end{align} 
Here for task 1 we are using the model specified in \eqref{eqtvb} as the underlying model, of which the parameters are estimated using the  most recent 24 months (from 01/2009 to 12/2010) of hourly load and temperature. The 10 years (2001-2010) of weather history is used to generate 10 weather scenarios. In total, we are getting 10 load forecasts for each hour in 01/2011. We compute the required 99 quantiles based on these 10 forecasts using the empirical distribution function.  
Similarly, we generate the 99 quantiles for the other 11 months of 2011. For instance, when forecasting the load of 05/2011, the 24 months from 05/2009 to 04/2011 of hourly load and temperature is used for parameter estimation.

\subsection{Recency model}

The underlying model for the second benchmark is given as:
\begin{align}
\label{eqrecency}
Y_{\LL,t} 
= \beta_0 &+ \beta_1 \text{MoY}(t) + \beta_2 \text{DoW}(t) + \beta_3 \text{HoD}(t) + \beta_4 \text{DoW}(t)\text{HoD}(t)   \nonumber\\
& + f(Y_{\TT,t}) + \sum_{j\in \JJ} f(\wtilde{Y}_{\TT,t,j}) + \sum_{k\in \KK} f(Y_{\TT, t-k}) + \epsilon_t,
\end{align}
where $f$ is as in \eqref{eq:SisterModel:2} and the daily moving average temperature of the $j$-th day $\wtilde{Y}_{\TT,t,j}$ is defined through
\begin{equation}
\label{eqrecency2}
\wtilde{Y}_{\TT,t,j} = \frac{1}{24} \sum_{h=24j-23}^{24j} \wtilde{Y}_{\TT, t-h}.
\end{equation}
The sets $\JJ$ and $\KK$ in equation \eqref{eqrecency} are given by $\JJ = \{1, \ldots, J\}$ and $\KK = \{1, \ldots, K\}$
for $J> 0$ and $K>0$; they are empty if $J=0$ and $K=0$. Note that for $(J, K) = (0,0)$ we receive the \textit{Vanilla} in \eqref{eqtvb}. 
The 'average-lag' pair $(J, K)$ needs to be identified before 
the \textit{Recency} model could be applied to generate forecast for the target month. 
Since the load pattern against temperature varies each year, the optimal pair selected correspondingly changes every year. To identify the optimal pair for the year $i$, we use the data of year $(i-3)$ and $(i-2)$ as training, the data of year $(i-1)$ as validation. 
The pair resulting in the lowest mean absolute percentage error (MAPE) in validation period will be selected and then the corresponding \textit{Recency} model will be  applied to forecast the year $i$. 
We search for the optimal $(J, K)$ on the grid $\{0, \ldots, 7\}\times \{0, \ldots, 48\}$.
With this method, the optimal pair identified for the year of 2011 is $(2,10)$ for the GEFCom2014-L data.


In the GEFCom2014-E case study, the target years are from 2010 to 2014. The optimal pairs identified are listed in Table \ref{tab_aa}. After identifying the optimal pairs of $(J,K)$, we follow the same steps as for the first benchmark discussed in Section 4.1, including two years of hourly loads and temperatures for parameter estimation and an empirical distribution function for extrapolating the 99 quantiles. But we use a \textit{Recency} model as the underlying model to do forecasting, instead of the vanilla model.  When creating weather scenarios, we use 6 years (2004-2009) weather data for the target year of 2010, 7 years (2004-2010) for 2011, 8 years (2004-2011) for 2012, 9 years (2004-2012) for 2013 and 10 years (2004-2013) for 2014.

\begin{table}[tbh]
\centering
\begin{tabular}{rrrrrr}
 Year & 2010 & 2011 & 2012 & 2013 & 2014  \\ 
  \hline 
$J$ & 1 & 1 & 1 & 1  & 0 \\  
$K$ & 9 & 0 & 8 & 13 & 13\\ 
\end{tabular}
\caption{The optimal pairs of $(J,K)$ for the years from 2010 to 2014 in GEFCom2014-E}
\label{tab_aa}
\end{table}

To keep the benchmarks simple and easy to reproduce, neither underlying models incorperate any other special treatments such as weather station selection, data cleansing, weekend and holiday effect modeling, or forecast combination.

\section{Empirical Results and Discussion}

We evaluate the forecasting performance by the overall mean pinball loss function of the 99 percentiles.
For more details on the pinball loss function and evaluation methods used in GEFCom2014-L, see \cite{hong2015probabilistic}.

\subsection{GEFCom2014-L results}

\begin{figure}[hbt!]
\centering
 \includegraphics[width=.99\textwidth, height=.34\textwidth]{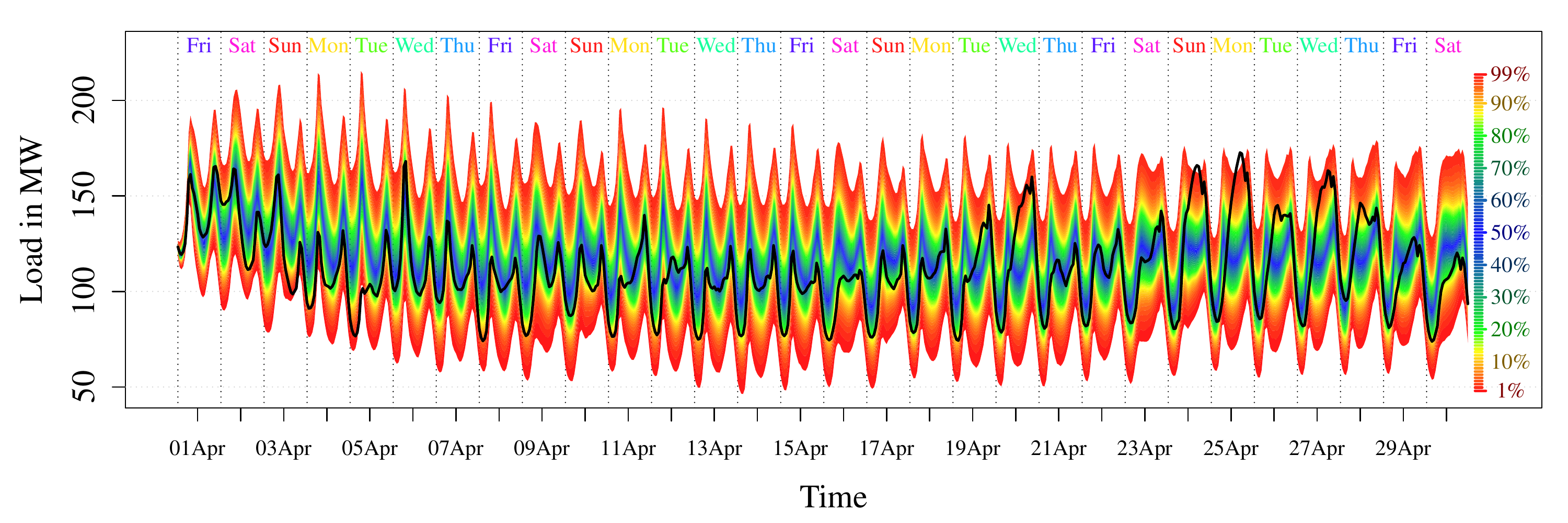} 
\caption{April forecast of the GEFCom2014-L data with corresponding legend and observed values (black line).}
 \label{fig_forc_example}
\end{figure}

 As an illustrative example, the predicted 99 quantiles for the April 2011 task are given in Figure 
 \ref{fig_forc_example}. 
We observe that the daily and weekly seasonal 
behaviors are well captured. 
Furthermore, the prediction intervals get wider with increasing forecasting horizon as expected. 

The pinball scores of the proposed model (Lasso) and the two benchmarks are given in 
Table \ref{tab_pb2}. We also list Bidong Liu's original GEFCom2014-L scores in the last column under BL. The main factors resulting in the difference between the two benchmarks and BL include the length of training data and the extrapolation method. 
In GEFCom2014-L, Bidong Liu implemented the scenario based method as described in section 4  for months 2 to 12, but not month 1. 
For parameter estimation, Bidong Liu used 5 years of historical data for most of the tasks during GEFCom2014-L. 
In addition, the required quantiles were generated by linear extrapolation.
For illustration purpose, we also list  
the pinball scores from the Vanilla benchmark estimated using 5 years of data in Table \ref{tab_pb2} under \textit{Vanilla}-5Y. 

We observe that the proposed lasso estimation method outperforms the two benchmarks, i.e.\textit{Vanilla} and \textit{Recency} in 9 and 8 months out of 12. 
The reductions on the 12-month average pinball score are 6.4\% and 7.6\% comparing with the 
\textit{Recency} and \textit{Vanilla}, respectively.
Although BL ranked top 8 in GEFCom2014-L, its average pinball score is higher than all the other four methods. 
The average pinball score of \textit{Vanilla}-5Y(8.32) is high than \textit{Vanilla}(8.05), which reveals the necessity of selecting the right length of the training data.

\begin{table}[tbh]
\centering
\begin{tabular}{rrrrrr}
 Month & Lasso &   \textit{Vanilla} &  \textit{Recency} &BL  &  \textit{Vanilla}-5Y \\   \hline \hline
 1 & \textbf{9.88}  & 11.94 & 12.13 & 16.42  &  11.78	\\ \hline  
 2 & \textbf{9.54} & 10.95  & 10.57 & 11.87 & 11.24 	\\ \hline  
 3 & \textbf{7.97}  &  8.57  & 8.38 & 9.37 &  8.70 \\ \hline  
 4 & 4.89  &  5.05  & \textbf{4.80}  & 5.62  &  5.67\\ \hline  
 5 & \textbf{5.96}  &  7.37  & 7.11 & 7.74 &  7.98 \\ \hline  
 6 & \textbf{5.86}  &  6.75  & 7.35  & 6.55 & 6.48 \\ \hline  
 7 & \textbf{7.66}  & 9.60 & 9.38 & 9.14  & 9.08 \\ \hline  
 8 & \textbf{10.70}  & 11.21   &11.30 & 11.35 & 11.36 \\ \hline  
 9 & 6.28  & 5.81  & \textbf{5.65} & 6.51 &  6.19 \\ \hline  
 10 & 5.20  & 3.53  & \textbf{3.40} & 4.80 &  4.53 \\ \hline  
 11 & 6.38  & 6.06  &\textbf{5.93} & 6.97 &  6.50	\\ \hline  
 12  & \textbf{8.99} & 9.74   & 9.45 &	10.89 & 10.29 \\ \hline   \hline
Average &  \textbf{7.44} & 8.05   & 7.95 & 8.94 & 8.32
\end{tabular}
\caption{ Overall pinball scores for the GEFCom2014-L data}
\label{tab_pb2}
\end{table}


\subsection{GEFCom2014-E results}

The pinball scores of the proposed method (Lasso) and the two benchmarks in GEFCom2014-E case study are given in 
Tables \ref{tab_pb1}.
We also provide the original scores of  Florian Ziel (FZ) in the GEFCom2014-E.
The FZ scores slightly differ from the Lasso, because
the long term trend components ($\GG_5$) were added to the time-varying parameters of Lasso. 
For FZ, no long term modeling was considered, but for the years 2012 and 2013
a manual long-term effect adjustment was done. Additionally, the list of considered holidays 
was extended by some bridging holidays, such as Christmas Eve (24 Dec), Boxing Day (26 Dec) and New Years Eve (31 Dec).

Similarly to the GEFCom2014-L results, the lasso outperforms the two benchmarks in 4 out of 5 years.
The average reductions of the pinball score in comparison with the
 \textit{Recency} and the  \textit{Vanilla} are 11.9\% and 15.6\%, respectively.
%


\begin{table}[tbh]
\centering
\begin{tabular}{rrrrr}
 Year & Lasso & FZ &\textit{Vanilla}  &  \textit{Recency} \\   \hline \hline
2010 & 59.01 & \textbf{58.02} & 85.03 & 80.76	\\ \hline  
2011 & \textbf{49.74} & 54.50 & 59.54 & 56.77	\\ \hline  
2012 & 47.08 & \textbf{46.51} & 57.58 & 55.37	\\ \hline  
2013 & 62.53 & 63.71 & 62.59 & \textbf{60.62}	\\ \hline  
2014 & 55.00 & \textbf{52.25} & 59.16 & 56.82	\\ \hline  \hline 
Average & \textbf{54.69} & 55.00 & 64.78 & 62.07
\end{tabular}
\caption{ Overall pinball scores for the GEFCom2014-E data}
\label{tab_pb1}
\end{table}


\subsection{Discussion}

Even though the proposed methodology outperforms two credible benchmarks, we may further improve it from several aspects. 
One model assumption is the homoscedasticity of the residuals, but the residuals are heteroscedastic in practice. 
Usually we observe lower variation in night
and during low load seasons. The heteroscedasticity of residuals should be taken into account when designing the model.
\cite{ziel2015efficient} and \cite{zieliteratively} suggest an iteratively reweighted lasso approach incorporating the volatility of the residuals. Their results suggest a significant improvement of the forecasting results. 
It might help as well to apply normality assumption with group analysis as discussed by \cite{XieHong} or a block bootstrap method as used by  \cite{fan2012short}, 
to incorporate the remaining dependency structure in the residuals. 
Another issue is the tuning of the lasso itself. We simply considered the Bayesian information criterion, but 
other special cases of the generalized information criterion (GIC) might yield better forecasting performance. 
Lastly, for the GEFCom2014-L data, the treatment of the available temperature information might be improved.
For instance, the weather station selection methodology as proposed by \cite{hongwhitepu2015} might yield a
better incorporation of the temperature data.



\section{Summary and Conclusion}

We introduce a lasso estimation based methodology 
that can estimate parameters for a large pool of candidate variables to
 capture several distinct and well-known stylized facts in load forecasting. 
The proposed methodology ranked top 2 in GEFCom2014-E. Two empirical studies based on two recent probabilistic load forecasting competitions (GEFCom2014-L and GEFCom2013-E) demonstrate the superior competence of the proposed method over two credible benchmarks. 


%
%
%


\section{References}

\small
 \bibliographystyle{apalike}
 \bibliography{article}

\end{document}